\documentclass[sigconf]{acmart}
\usepackage{amsmath}

\AtBeginDocument{%
  \providecommand\BibTeX{{%
    \normalfont B\kern-0.5em{\scshape i\kern-0.25em b}\kern-0.8em\TeX}}}

\begin{document}

\title{Text mining policy: Classifying forest and landscape restoration policy agenda with neural information retrieval}

\author{John Brandt}
\email{john.brandt@wri.org}
\affiliation{%
  \institution{World Resources Institute}
  \city{Washington}
  \state{D.C.}
}


\begin{abstract}
  Dozens of countries have committed to restoring the ecological functionality of 350 million hectares of land by 2030. In order to achieve such wide-scale implementation of restoration, the values and priorities of multi-sectoral stakeholders must be aligned and integrated with national level commitments and other development agenda. Although misalignment across scales of policy and between stakeholders are well known barriers to implementing restoration, fast-paced policy making in multi-stakeholder environments complicates the monitoring and analysis of governance and policy. In this work, we assess the potential of machine learning to identify restoration policy agenda across diverse policy documents. An unsupervised neural information retrieval architecture is introduced that leverages transfer learning and word embeddings to create high-dimensional representations of paragraphs. Policy agenda labels are recast as information retrieval queries in order to classify policies with a cosine similarity threshold between paragraphs and query embeddings. This approach achieves a 0.83 F1-score measured across 14 policy agenda in 31 policy documents in Malawi, Kenya, and Rwanda, indicating that automated text mining can provide reliable, generalizable, and efficient analyses of restoration policy.
\end{abstract}



\maketitle

\section{Introduction}

Restoration, which describes means to improving the ecosystem function of degraded landscapes, encompasses a variety of land use interventions such as agroforestry, the establishment of buffer zones, native species planting, natural regeneration, reforestation, and silviculture, among others \cite{chazdonetal}. Although the primary goals of restoration may be to enhance resilience and adaptation to climate change, restoration can also increase economic livelihoods and improve public health \cite{cobenefit}. The social and political contexts within which restoration strategies are situated are especially important because restoration involves trade offs between decisions that affect the primary income source of millions of people and businesses \cite{dudley_mansourian_vallauri_2005}. However, the political context of restoration is complicated by its broad nature, being regulated by natural resources as well as 
economic, forestry, and agricultural government bodies. In order to facilitate efficient implementation of restoration, each of these political stakeholders must coalesce at the landscape scale while both supporting community participation and maintaining alignment with national and international development agenda \cite{Holl455}. 

Successful restoration strategies depend upon evidence-driven analyses of restoration policy that identify and integrate the agenda, values, and needs of individual stakeholders to strike a balance of livelihoods, agricultural production, and conservation with other development agenda \cite{TRD} \cite{doi:10.1111/rec.12339}. This complicated multi-stakeholder decision making process presents a multitude of opportunities for both conflict and synergy \cite{GORG2007954}. For instance, Rwanda's agricultural intensification policies conflicted with the agroforestry approach promoted by the Ministry of Natural Resources \cite{VANOOSTEN201863}. Other interdisciplinary policy issues, such as land tenure and property rights, are central to restoration implementation but have yet to be well integrated in monitoring frameworks \cite{MCLAIN2018}. Network governance, an informal arrangement where stakeholders regularly participate in knowledge and information sharing, has shown success as an approach to landscape-level governance by addressing policy conflicts among fragmented governance structures \cite{networkgovernance}. The potential for conflicting agenda in multi-stakeholder environments and the lack of mainstreaming of analyses of political and social economies in restoration monitoring identify a need for quantitative assessments of policy agenda to design successful restoration strategies and promote collaborative planning efforts \cite{TRD} \cite{Holl455}.

Within the field of political science, content analysis methods such as topic classification allow researchers to identify patterns and changes in political agenda over time and across geographies \cite{Hillard07anactive}. Text mining, which uses supervised and unsupervised natural language processing and machine learning methods for tasks such as classifying agenda topics in policy, has been shown to increase both the effectiveness and efficiency of policy making \cite{grimmer_stewart_2013}. Across the public health, homeland security, and privacy and financial regulations sectors, text mining has reduced data collection and information processing time while mitigating the problem of bounded rationality \cite{Ngai2016ARO} \cite{requirementsanalysis}. Supervised methods, such as conditional random field models and decision trees, have been applied to problems ranging from identifying business processes in corporate policy documents \cite{Li2010} to classifying high-level policy agenda \cite{Karan2016AnalysisOP}. Because supervised methods require large amounts of training data in order to effectively generalize to new contexts, unsupervised methods are often used when labeled training data is limited or when the goal is to discover patterns in the data \cite{grimmer_stewart_2013}. Unsupervised methods, such as topic modeling and embedding clustering, have been utilized in applications including identifying shifts in policy agenda in the European Parliament \cite{Greene:2015:UPA:2786451.2786464} and classifying topics in public government petitions \cite{publiccomments}. 

Although much research has considered the applications of text mining to policy research, there remain a number of challenges to their widespread implementation. One such challenge is that specific policy agenda are often sparsely distributed within policy documents. Each policy document is also likely to incorporate many different agenda, which may be difficult to disambiguate due to either similarities between agendas or differences in prose and diction between documents. While policy analysis is often performed with a prior knowledge of class labels, supervised classification methods are not well suited to tasks where an individual sentence can govern the classification of documents which may contain thousands of sentences \cite{SUN2009191}. Classification at the page, paragraph, or sentence level may mitigate this issue, but increasing the granularity of classification requires a more strenuous labeling process and introduces issues of class imbalance. On the other hand, unsupervised clustering methods do not allow the researcher to establish pre-determined topics, and results often times may not suit the needs of the researcher \cite{grimmer_stewart_2013}. Information retrieval, which ranks documents based on their relevance to a query, is not commonly used for classification but does not suffer from issues of class imbalance, multi-class labeling, or large training data requirements. This paper addresses these issues facing text mining in policy analysis by introducing a neural information retrieval approach to classifying policy agenda with word embeddings.

\section{Methods}
\label{Methods}

\begin{figure*}[ht]
\vskip 0.2in
\begin{center}
\centerline{\includegraphics[width=14cm]{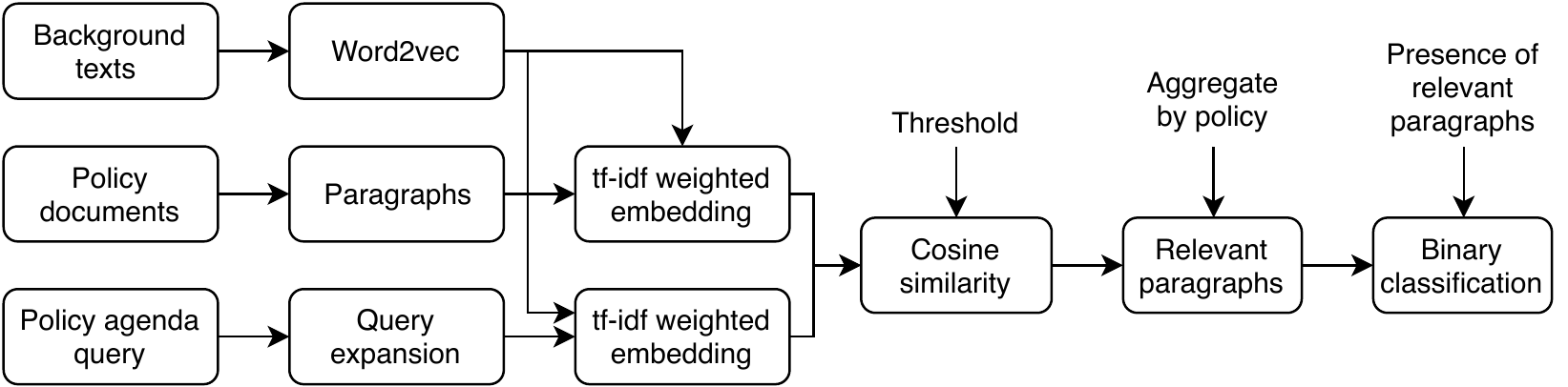}}
\caption{Overview of classification architecture applied to each of the fourteen studied policy agenda.}
\label{model-figure}
\end{center}
\vskip -0.2in
\end{figure*}

To test the applicability of information retrieval with word embeddings to classifying restoration policy agenda, we focus on 31 policy documents from Kenya, Rwanda, and Malawi concerning agriculture, national development agenda, economic development, forestry, land, and natural resources (Table \ref{policies-table}). Labeled response data were generated by policy experts who identified the binary presence of 14 policy agenda and specific restoration interventions based on a close reading of each policy text (Table \ref{model-table}). For each labeled response, we generate classifications indicating the presence or absence of a policy agenda by querying paragraph-level feature vectors generated from the word embeddings (Figure \ref{model-figure}).

\begin{table}[h!]
\vskip -0.1in
\caption{Categories of policy documents. General refers to any national, sector agnostic policy such as the constitution or national development plans.}
\label{policies-table}
\vskip 0.15in
\begin{center}
\begin{small}
\begin{sc}
\begin{tabular}{lcccr}
\toprule
Sector/Ministry & Kenya & Malawi & Rwanda \\
\midrule
Agriculture & 3 & 0 & 0\\
General & 3 & 2 & 2\\
Economy & 1 & 2 & 1\\
Forestry & 2 & 5 & 3\\
Land & 2 & 1 & 0\\
Natural Resources & 1 & 2 & 1\\
\midrule
\textbf{Total} & \textbf{12} & \textbf{12} & \textbf{7} \\
\bottomrule
\end{tabular}
\end{sc}
\end{small}
\end{center}
\vskip -0.1in
\end{table}

\begin{table}[h!]
\vskip -0.1in
\caption{Restoration policy agenda categories grouped by their relevance to natural and ecological outcomes and social and governance outcomes.}
\label{model-table}
\vskip 0.15in
\begin{center}
\begin{small}
\begin{sc}
\begin{tabular}{ll}
\toprule
Natural \& Ecological & Social \& Governance \\
\midrule
Maintaining \# of trees & Economic benefits \\
Increasing \# of trees & Health benefits \\
Silviculture & Benefit sharing \\
Agroforestry & Land ownership \\
Soil erosion & Land use rights \\
Forest protection & Local participation \\
Buffer zones & Timber product mgmt. \\
\bottomrule
\end{tabular}
\end{sc}
\end{small}
\end{center}
\vskip -0.1in
\end{table}

Policy documents were converted to plain text with optical character recognition (OCR) and cleaned by removing page numbers, citations, and section headers with regular expression. Spelling mistakes introduced in the OCR process were corrected with the  Hunspell library in a context-aware manner by choosing the most likely spelling according to each document's word distribution. Documents were subsetted into sentences and subsequently into paragraphs. 

To ensure that our approach is able to generalize to restoration concepts that are not found verbatim within the policy documents themselves, we leverage a corpus of background documents for model training, and apply pre-trained models to the study documents with transfer learning. A total of 1,512 restoration-relevant documents were identified by searching for open access journal articles about restoration, sector agnostic environmental policies, reports written by development organizations about climate change or the environment, and policy analyses written by nonprofits and academics about restoration. Word embeddings with 300 dimensions were generated for unigrams, bigrams, and trigrams in the background corpus by using the Skip-gram method of \cite{NIPS2013_5021}, commonly referred to as Word2vec. Presented results use a window size of 12, a minimum count of 15, and a negative sample size of 15.

The Skip-gram approach to Word2vec uses a surrogate neural network to find representations of words that are useful for predicting context words. Word embeddings are given by the hidden layer of a neural network taking as input individual words and outputting probabilities over the entire vocabulary that each word is located within the context of the input. Given a sequence of words, $w_{1...T}$, Skip-gram seeks to maximize the average log probability given in Equation 1, where $c$ is the number of context words considered, and $p(w_{t+j}|w_t)$ is defined with the hierarchical softmax function. This approach outputs similar word embeddings for words that appear in similar contexts. 

\begin{equation}
    \frac{1}{T}\sum\limits_{t=1}^T \sum\limits_{-c\leq j\leq c,j\neq0} \log p(w_{t+j}|w_t)
\end{equation}

Embeddings were generated for separate paragraphs within the study policy documents by weighting the word embeddings of constituent words, calculated using the background corpus, by term frequency-inverse document frequency (tf-idf, Equation 2).

\begin{equation}
\text{tf-idf} = \frac{f_{t,d}}{\sum f_{t',d}}\log\frac{N}{f_{d,t}}
\end{equation}

Weighting by tf-idf increases the importance of words that are unique to a paragraph or sentence. Information retrieval methods were used as a proxy for classification by retrieving paragraphs with high similarity to input queries for each policy agenda or restoration intervention. For multi-word queries, we calculate the tf-idf reweighted vector of the constituent embeddings. Specifically, policy documents are classified as relevant to a given response if at least one paragraph in the document is within a cosine similarity threshold of the respective query. Cosine similarity (Equation 3) measures the similarity between two vectors, with zero and one representing no and complete similarity, respectively.

\begin{equation}
    \theta(A,B) = \frac{A\cdot B}{||A||\cdot||B||}
\end{equation}

Queries and cosine similarity thresholds were determined in an interactive and iterative manner. The fifty closest related words and n-grams to the tf-idf weighted embedding vector of words exactly matching the target labels were qualitatively evaluated for relevance to the target label. Closely related words were iteratively added or removed from the query, until the query either reached five words, or the addition or removal of words no longer changed the related words and n-grams. We selected separate cosine similarity thresholds for each label to deal with  differences in the generality of topics. For example, we expect the embedding space containing words related to the target label "health benefits" to be larger than that of "soil fertility" because there are more ways to discuss the former than the latter. Recognizing that this generality is subjective, we iteratively decrease the cosine similarity threshold from 0.55 in steps of 0.01 until sentences not relevant to the query are returned, measured by a randomized reading of the retrieved paragraphs with the lowest cosine similarity. This approach can be conceptualized as separate bivariate clustering of the embedding for each topic where one cluster is centered around the query embedding point and the researcher instills a sort of weak supervision over the classifications by explicitly choosing the similarity threshold within which class membership is assigned.

Policy documents containing at least one paragraph following the above-described information retrieval process were classified as relevant to the respective policy agenda or topic. We ensure the unsupervised nature of this approach by only comparing results with human-generated labels for the results of the selected queries and thresholds. The performance of the information retrieval classification was evaluated with precision, accuracy, recall, and F1-score based upon the human-generated labels. Summary reports containing the relevant paragraphs, labeled by policy document and page number, were generated for each query to ensure replicability and to analyze consistency, reliability, and relevance of results.

\section{Results}
\label{Results}

Measured across 31 policy documents in Malawi, Kenya, and Rwanda, the present approach of leveraging information retrieval queries as a classification label achieves an average of 0.83 F1-score on classifying the binary presence of 14 policy agenda (Table \ref{res-table}). The methodology performed best on topics with a very narrow scope and little overlap with other agenda, such as the establishment of buffer zones, and worst on topics that are broad in nature, subjective, and have potential to overlap with other agenda, such as land use rights and forest protection. Overall, we report similar metrics across the three countries indicating that this approach is able to generalize to new contexts. Specifically, the F1-scores for Kenya, Malawi, and Rwanda were 0.85, 0.80, and 0.82, respectively.

\begin{table*}[h]
\caption{Example sub-passages classified and their assigned agenda labels based on the information retrieval process.}
\label{examples-table}
\vskip 0.15in
\begin{center}
\begin{small}
\begin{tabular}{ll}
\toprule
Agenda & Passage \\
\midrule
Maintaining \# of trees &  Harvesting of trees shall be done ... to maintain a 10\% tree cover.\\
Increasing \# of trees &  Shall identify land ... at the risk of land degradation and institute measures necessary for \\ 
& \hspace{0.5cm} ensuring its conservation including planting of trees. \\
Economic benefits &  Lead to improved rural incomes. \\
Local participation & De-concentrate responsibility for planning and implementing management to grassroots levels. \\
Agroforestry & Trees on croplands stabilize the soil and improve soil fertility. \\
Buffer zone & Such as trees planted in strips on boundaries and between fields.  \\
Timber product mgmt. &  Plantation forests ... shall be managed on a sustainable basis for the production of wood.\\
\bottomrule
\end{tabular}
\end{small}
\end{center}
\end{table*}

\begin{table}[h!]
\vskip -0.1in
\caption{Accuracy, precision, recall, and F1-score  for policy agenda classification across 31 policy documents in Malawi (n=12), Kenya (n=12), and Rwanda (n=7).}
\label{res-table}
\vskip 0.15in
\begin{center}
\begin{small}
\begin{sc}
\begin{tabular}{lcccr}
\toprule
Agenda & Accuracy & Precision & Recall & F1 \\
\midrule
Maintaining \# trees &  0.90 & 1.00 & 0.90 & 0.97 \\
Increasing \# trees & 0.71 & 0.75 & 0.78 & 0.77 \\
Economic benefits & 0.84 & 0.89 & 0.84 & 0.86 \\
Health benefits & 0.74 & 0.88 & 0.77 & 0.82 \\
Benefit sharing & 0.88 &  0.92 & 0.55 & 0.75 \\
Land ownership & 0.87 &  0.83 & 0.89 & 0.85\\
Land use rights & 0.65 &  0.66 & 0.78 & 0.73\\
Local participation & 0.90 & 0.92 & 0.96 & 0.94 \\
Silviculture & 0.87  &  0.89 & 0.82 & 0.87 \\
Agroforestry & 0.87  &  0.94 & 0.78 & 0.90 \\
Soil erosion &  0.87 & 0.95 & 0.86 & 0.89 \\
Forest protection & 0.68 & 0.48 & 0.60 & 0.54 \\
Buffer zone & 0.94 & 0.87 & 1.00 & 0.92 \\
Timber product mgmt. & 0.85 & 0.80 & 0.90 & 0.86 \\
\midrule
\textbf{Average} & \textbf{0.83} & \textbf{0.82} & \textbf{0.84} & \textbf{0.83} \\
\bottomrule
\end{tabular}
\end{sc}
\end{small}
\end{center}
\vskip -0.1in
\end{table}

Table \ref{examples-table} shows example sub-passages returned by the neural information retrieval process. As can be seen, the Word2vec model trained on a background corpus of environmental policy documents effectively learns to associate relevant phrases with their correct agenda label with little or no direct diction overlap. For instance, the methodology accurately predicts "trees planted in strips on boundaries" as constituting buffer zones, and "de-concentrating responsibility... to grassroots levels" as local participation in government. 

Query expansion was only needed in two of the fourteen agenda contexts, with "local participation" being expanded to include "community involvement", and "increasing trees" being expanded to include "afforestation". The default cosine similarity threshold of 0.55 generally performed well, and selected thresholds varied between 0.48 and 0.55. However, fine tuning of these parameters was useful for establishing a boundary for broad and complicated agenda topics. False positives primarily occurred in agenda topics of land use rights, increasing the number of trees, and forest protection. A manual inspection of false positives for these agenda found that land tenure was often conflated with land use rights, conservation and sustainable forest management were often incorrectly classified as increasing the number of trees, and issues of forest protection were often mentioned in the context of other agenda without being a stated focus of the policy. 

\section{Conclusion}
\label{Conclusion}

In this paper, we have proposed a novel method of classifying policy agenda by leveraging neural information retrieval to address issues of class imbalance, limited availability of labeled training data, and the importance of single sentences within much longer documents. Evaluated on dozens of policy documents across multiple ministries and countries, the results suggest that the method is generalizable and robust to changes in diction and prose between governmental contexts. In comparison to traditional policy analysis, the presented method is more scalable, replicable, and objective. For each agenda query, the methodology outputs a formatted report with metadata, text extractions, and page references for each relevant passage, promoting replicability and enforcing objectivity of the classification decisions. Additionally, the incorporation of human supervision in the query expansion and threshold selection stage allows the researcher to manually adjust the methodology for fine tuning between domains.

With regards to restoration, the results show that neural information retrieval can accurately extract and classify diverse policy agenda from a broad range of policy documents. This brings potentially significant implications for governance issues in restoration. The need for information sharing and communication across ministries of government and the gap between international commitments and local-level priorities are well known barriers to successful restoration strategies \cite{silo}. Transparency in policy agenda between stakeholders is especially important in restoration because it requires an integrated landscape approach to governance \cite{Adhikari:2016:1465-5489:265}.  These results suggest that text mining can effectively extract and classify important policy agenda from thousands of pages of policy documents to foster knowledge sharing between stakeholders. Accurately and efficiently extracting and classifying policy agenda opens the door to applications such as identifying conflicting agenda between ministries or scales of government or identifying where synergies exist between policies to promote better alignment of restoration strategies.

Another major hurdle to restoration implementation involves mainstreaming restoration in both social and environmental policy. While restoration provides economic and social benefits and trade-offs, they are often not properly integrated in health and economic development policy \cite{Menz526}. In order to achieve national-scale implementation of restoration and sustainable land use agenda, integrated and multi-sectoral analyses that prioritize interventions that balance societal, economic, and environmental trade offs are required \cite{Gao2017}. This paper's results indicate that text mining can be used to identify gaps in policy, and to subset relevant passages of policies to quickly summarize how the social, economic, and environmental trade-offs of restoration are framed and where they are missing from the conversation. 


While the present research establishes the utility of analyzing restoration policy with neural information retrieval, a number of challenges remain that may be addressed by future research. Two primary limitations of this study are its focus on national-level policies and on policy agenda rather than specific incentives and disincentives for restoration. As such, the next steps for research include analyzing local-level policies and analyzing the sentiment around policy agenda and specific financial incentives. Additionally, text mining models based on Word2vec have difficulty disambiguating word meanings in different contexts and must be trained specifically on texts from the application domain \cite{elmo}. This implies that models such as the one presented in this paper should be trained individually for each policy domain to achieve maximum interpretability. Although this paper demonstrated fairly accurate results on cross-domain applications, models which can learn to disambiguate word meanings based on the domain context would likely fare better. Word embedding based approaches also struggle to differentiate synonyms and antonyms, and require word-level features to be aggregated to the sentence or paragraph level \cite{8023397}. Although the approach presented in this paper effectively identifies passages pertaining to specific policy agenda, these drawbacks contribute to the method's inability to differentiate between ambitions, discussions of agenda, and actual implementation strategies and targets. Recent embedding approaches which learn word sense disambiguation, such as ELMo embeddings \cite{elmo}, or explicitly model variable length passages, such as Doc2Vec \cite{doc2vec} or skip-thoughts \cite{skipthoughts}, may be useful in extending the generalizability and specificity of this approach. Finally, the utility of this approach may be extended by training light-weight classification algorithms on top of the results to automate query expansion and cosine similarity threshold selection or to classify the sentiment and subjectivity of passages.



\begin{acks}
Thanks to Kathleen Buckingham for providing expert advice on restoration policy and agenda, Sabin Ray for feedback on methodology, and Kai Qi for generating the labeled training data.
\end{acks}

\bibliographystyle{ACM-Reference-Format}
\bibliography{bib}

\appendix

\end{document}